\documentclass{article}
\usepackage{epsfig}

\newcommand{\mean}[1]{\langle #1 \rangle}

\begin{document}
\title{Social Structure and Opinion Formation}
\author{Fang Wu\thanks{Applied Physics Department, Stanford University, Stanford, CA 94305}
~and Bernardo A. Huberman\thanks{HP Labs, Palo Alto, CA 94304}}
\maketitle

\begin{abstract}
We present a dynamical theory of opinion formation that
takes explicitly into account the structure of the social network
in which individuals are embedded. The theory predicts the
evolution of a set of opinions through the social network and
establishes the existence of a martingale property, i.e.~the
expected weighted fraction of the population that holds a given
opinion is constant in time. Most importantly, this weighted
fraction is not either zero or one, but corresponds to a
non-trivial distribution of opinions in the long time limit. This
coexistence of opinions within a social network is in agreement
with the often observed locality effect, in which an opinion or a
fad is localized to given groups without infecting the
\emph{whole} society. We verified these predictions as well as
others concerning the fragility of opinions and the importance of
highly connected individuals by computer experiments on scale-free networks.
\end{abstract}

\section{Introduction}

Most people hold opinions about a myriad topics, from politics and
entertainment to health and the lives of others. These opinions
can be either the result of serious reflection or, as is often the
case when information is hard to process or access, formed through
interactions with others that hold views on given issues. This
reliance on others to form opinions lies at the heart of
advertising through social cues, efforts to make people aware of
societal and health related issues, fads that sweep social groups
and organizations, and attempts at capturing the votes and minds
of people in election years.

Because of our dependence on other's to shape our views of the
world, an understanding of opinion formation requires an
examination of the interplay between the structure of the social
network in which individuals are embedded and the interactions
that take place within it. This explains the vast efforts that
both the commercial and the public sectors devote to uncovering
such interplay and the mechanisms they deploy to affect the
formation of favorable and unfavorable opinions about any
imaginable topic. More recently, the emergence of email and global
access to information through the web has started the change the
discourse in civil society  \cite{david, margolis, rash,
rheingold, wilheim}. and made it even easier to propagate points
of view and misleading facts through vast numbers of people; views
which are surprisingly accepted and transmitted on to others
without much critical examination.

In the academic arena there exist several models of opinion
formation that take into account some factors while leaving out
others (for one the earliest see \cite {french}). In economics,
information cascades were proposed to explain uniformity in social
behavior \cite{bikhchandani1992, bikhchandani1998}, as well as its
fragility. This approach assumes that there is a linear sequence
of Bayesian individuals that can observe the choices of others in
front of them before making their own decisions as to which
opinion to choose. Besides the notorious problems with assuming
Bayesian decision makers \cite{camerer} this theory makes
unrealistic assumptions, such as assuming a sequence of
synchronous decisions that does not take into account the social
network, or locality, of contacts that people have. More
problematic is the prediction that given an initial set of
possible opinions, the information cascade will lead to one
opinion eventually becoming pervasive, which contradicts the
common observation that conformity throughout a society tends to
be localized in subgroups rather than widespread.

Other approaches to opinion formation have dealt with either
theoretical models or computer simulations. On the theory side a
number of dynamical models that have been proposed are based on
analogies to magnetic systems placed either on a continuum or on a
two dimensional lattices \cite{hegselmann} or on a discrete
many-state space \cite{galam1997, galam2000, dodds}. While none of
them takes into account the social structure in shaping opinion
formation, on the side of computer simulations, models that have a
continuum of possible opinions or very large number of opinions
can sometimes yield asymptotic states that are non-uniform
\cite{deffuant, hegselmann, stauffer2003, stauffer2004,
fortunato}, partly due to the many choices of opinions. However,
all models with a binary choice of opinions do lead to widespread
dominance \cite{galam1998, galam2000, sznajd-weron, galam2002,
laguna, tessone}, once again in disagreement with observations.

In this paper we propose a theory of opinion formation that
explicitly takes into account the structure of the social network
in which individuals are embedded. The theory assumes asynchronous
choices by individuals among two or three opinions and it predicts
the time evolution of the set of opinions from any arbitrary
initial condition. We show that under very general conditions a
martingale property ensues, i.e. the expected weighted fraction of
the population that holds a given opinion is constant in time. By
weighted fraction we mean the fraction of individuals holding a
given opinion, averaged over their social connectivity (degree).
Most importantly, this weighted fraction is not either zero or
one, but corresponds to a non-trivial distribution in the long
time limit. This coexistence of opinions within a social network
is in agreement with the often observed locality effect, in which
an opinion or a fad is localized to given groups without infecting
the \emph{whole} society.

Our theory further predicts that a relatively small number of
individuals with high social ranks can have a larger effect on
opinion formation than individuals with low rank. By high rank we
mean people with a large number of social connections. This
explains naturally the fragility phenomenon, whereby an opinion
that seems to be held by a rather large group of people can become
nearly extinct in a very short time, a mechanism that is at the
heart of fads.

These predictions were verified by computer experiments and
extended to the case when some individuals hold fixed opinions
throughout the dynamical process. Furthermore, we dealt with the
case of information asymmetries, which are characterized by the
fact that some individuals are often influenced by other people's
opinions while being unable to reciprocate and change their
counterpart's views.

In the following sections we describe the dynamical model and
proceed to solve it analytically. We then extend it to several
interesting cases (fixed opinions and information asymmetries) and
then present the results of computer simulations that confirm the
theoretical predictions. A concluding section summarizes our
results and discusses their implications to opinion formation and
possible future research.

\section{Two opinions within a social network}

\subsection{Description of the model}

In our theory we represent a social network as a connected random
graph with a certain degree distribution $p_k$. The nodes of this
graph correspond to people and the edges represent their social
connection. We assume that the graph is entirely random except for
its degree distribution, which means that the degree of each node
is drawn independently from the distribution $p_k$, and any two
graphs with the same degree sequence are equally likely in the
sample space. This point is made clear in Section II of
\cite{newman}. In the following discussion we also assume that the
structure of the social network changes over time scales that are
much slower than opinion formation, so that for all practical
purposes the graph structure can be considered static over the
time that opinions form.

We use the terms ``black'' and ``white'' to denote the binary
opinions available to each person, who is represented by a node.
A person (node) is either of the black or of the white opinion. We
then assume that starting from an initial color distribution,
people asynchronously update their opinions at a rate $\lambda$.
That is, during any time interval $dt$, each node updates its
color (makes a decision as to which opinion to hold) with
probability $\lambda dt$, based on the colors of its neighbors.
Specifically, if a given person or node has $b$ black neighbors
and $w$ white neighbors, then the probability that its new color
is going to be black is $b/(b+w)$. This is equivalent to assume
that each time a person randomly chooses one of its neighbors and
sets its new color to be the same as that neighbor. Note that when
we say that a person or node ``updates'', we do not mean
``changes''. It is completely possible that a node remains the
same color after the update.

With this opinion adoption mechanism in place and a given social
structure we now determine how opinions spread throughout the as a
function of time. As we show, which opinion (or color) will
prevail is not obvious, as well as how the ratio of black-to-white
changes with time?

We will first consider the case where once the opinion formation
starts, no new sources of opinions enter the social network. We
will then relax this assumption by allowing for new opinions to
enter into a social network as time evolves.

Throughout this paper we will use the following symbols and their
meaning, which are listed in Table \ref{tab_symbols}.

\begin{table}
\caption{\label{tab_symbols}Symbols and their meanings}
\begin{center}
\begin{tabular}{|c|l|}
\hline
$n$ & total number of nodes\\
$n_k$ & number of nodes with degree $k$\\
$p_k=n_k/n$ & the degree distribution\\
$m$ & total number of black nodes\\
$m_k$ & number of black nodes with degree $k$\\
$q=m/n$ & fraction of black nodes\\
$q_k=m_k/n_k$ & fraction of black nodes in all degree-$k$ nodes\\
\hline
\end{tabular}
\end{center}
\end{table}

\subsection{The dynamics of opinion formation}

Consider a specific update that happens at some time $t$. Let $A$
be the person or node that updates, and let $k$ be its degree.
Because all $n$ nodes update their colors asynchronously and
independently of each other at the same rate, everyone has the
same chance to be observed updating at time $t$. Thus the degree
distribution of $A$ is just the degree distribution of a randomly
chosen node, or $p_k$. During the update, $A$ randomly copies the
color from one of its neighbors, which we will call $B$. We
calculate the change of $m_k$ due to this specific update. There
are three cases:

\begin{enumerate}
\item $A$ is white and $B$ is black. $A$ updates its color to
black and consequently increases $m_k$ by 1. \item $A$ is black
and $B$ is white. $A$ updates its color to white and consequently
decreases $m_k$ by 1. \item $A$ and $B$ have the same color. In
this case $m_k$ does not change.
\end{enumerate}

Given $A$'s degree $k$, the probability that $A$ is black or white
before the update is simply $q_k$ or $1-q_k$ by definition. To
calculate the black probability of $B$ we need to know its degree
distribution first, which in our case is not $p_k$. This is
because $A$ being a randomly chosen node is more likely to be a
neighbor of a high degree node than a low degree node.
Specifically, the probability that $B$ has degree $j$ is
proportional to $jp_j$ \cite{feld}. Conditioning on the event that $B$ has degree
$j$, the black probability of $B$ is again simply $q_j$.

Thus, the probability that the update changes a degree-$k$ node
from white to black (case 1) is given by
\begin{equation}
P_{w\to b}(k) = p_k (1-q_k) \frac {\sum_j j p_j q_j} {\sum_j j p_j}.
\end{equation}
Similarly, the probability that the update changes a degree-$k$
node from black to white (case 2) is given by
\begin{equation}
P_{b\to w}(k) = p_k q_k \frac {\sum_j j p_j (1-q_j)} {\sum_j j p_j} = p_k q_k \left(1 - \frac {\sum_j j p_j q_j} {\sum_j j p_j} \right).
\end{equation}
If we define
\begin{equation}
\label{weighted_average} \mean q = \frac {\sum_j j p_j q_j} {\sum_j j p_j}
\end{equation}
to be a weighted average over all $q_k$'s, then the two
probabilities can be written as
\begin{equation}
\label{prob_w_b}P_{w\to b}(k) = p_k (1-q_k) \mean q,
\end{equation}
\begin{equation}
\label{prob_b_w}P_{b\to w}(k) = p_k q_k (1-\mean q).
\end{equation}
This gives us the increment of $m_k$ due to a particular update:
\begin{equation}
\Delta m_k = \left\{
\begin{array}{rl}
+1 & \mbox{with probability }p_k (1-q_k) \mean q\\
-1 & \mbox{with probability }p_k q_k (1-\mean q)\\
0  & \mbox{otherwise}
\end{array}
\right..
\end{equation}

Note that the updating process of the whole network (not just one
node) is a Poisson process of rate $n\lambda$. Hence the increment
of $m_k$ in a time interval $(t,t+dt)$ is given by
\begin{equation}
\Delta m_k = \left\{
\begin{array}{rl}
+1 & \mbox{with probability }n_k (1-q_k) \mean q \lambda dt\\
-1 & \mbox{with probability }n_k q_k (1-\mean q) \lambda dt\\
0  & \mbox{otherwise}
\end{array}
\right.,
\end{equation}
where we used the fact $n_k=np_k$.

We can now calculate the expectation and variance of the random
variable $\Delta m_k$. Its expectation is given by
\begin{equation}
E[\Delta m_k] = n_k (1-q_k) \mean q \lambda dt - n_k q_k (1-\mean q) \lambda dt = n_k (\mean q-q_k) \lambda dt.
\end{equation}
Its second moment is equal to
\begin{eqnarray}
E[(\Delta m_k)^2] &=& n_k (1-q_k) \mean q \lambda dt + n_k q_k (1-\mean q) \lambda dt\nonumber\\
&=& n_k (\mean q + q_k - 2\mean q q_k) \lambda dt.
\end{eqnarray}
Hence the variance is given by
\begin{eqnarray}
\mbox{var}[\Delta m_k] &=& E[(\Delta m_k)^2] - (E[\Delta m_k])^2 \nonumber\\
&=& n_k (\mean q + q_k - 2\mean q q_k) \lambda dt + o(dt)\nonumber\\
&=& n_k \sigma_k^2 \lambda dt + o(dt),
\end{eqnarray}
where
\begin{equation}
\sigma_k^2 \equiv \mean q + q_k - 2\mean q q_k.
\end{equation}

By definition, $q_k=m_k/n_k$, so we have (to $dt$ order)
\begin{equation}
\label{mean_delta_qk} E[\Delta q_k] = \frac 1{n_k} E[\Delta m_k] = (\mean q-q_k) \lambda dt,
\end{equation}
and
\begin{equation}
\label{var_delta_qk} \mbox{var}[\Delta q_k] = \frac 1{n_k^2} \mbox{var}[\Delta m_k] = \frac 1{n_k} \sigma_k^2 \lambda dt.
\end{equation}

The increment step of $\Delta q_k$ is $1/n_k$. When $n$ is large
this step is small, and Eq.~(\ref{mean_delta_qk}) and
(\ref{var_delta_qk}) can be approximated by a continuous process
described by the following stochastic differential equation
\begin{equation}
\label{equation_with_lambda} dq_k = (\mean q-q_k) \lambda dt + \frac 1{\sqrt{n_k}} \sigma_k \sqrt \lambda \,dB_t^{(k)},
\end{equation}
where $B_t^{(k)}$ are $k$ independent Brownian motions. From now on we redefine the time unit so that $\lambda=1$. Then Eq.~(\ref{equation_with_lambda}) becomes
\begin{equation}
\label{dynamics} dq_k = (\mean q-q_k) dt + \frac 1{\sqrt{n_k}} \sigma_k dB_t^{(k)}.
\end{equation}
which is the set of equations that governs the dynamics of the
social network.

\subsection{The solution}

\subsubsection{Martingale}

The quantities $q_k$ and $\mean q$ in Eq.~(\ref{dynamics}) are all
random variables, and $\sigma_k$ is nonlinear in $q_k$. As a
result Eq.~(\ref{dynamics}) is very hard to solve. However,
observe that if we take the weighted average (see
Eq.~(\ref{weighted_average})) of both sides of
Eq.~(\ref{dynamics}), we obtain
\begin{equation}
d\mean q = \mean{\frac 1{\sqrt{n_k}} \sigma_k dB_t^{(k)}},
\end{equation}
or
\begin{equation}
\mean{q(t)} = \int_0^t \mean{\frac 1{\sqrt{n_k}} \sigma_k
dB_s^{(k)}} = \left( \sum_k k p_k \right)^{-1} \sum_k \frac{k
p_k}{\sqrt{n_k}} \int_0^t \sigma_k dB_s^{(k)}.
\end{equation}
Because the right hand side does not include the $dt$ term, $\mean
q$ is a martingale. Thus its expectation value does not change
with time:
\begin{equation}
\label{eq_constant} E[\mean{q(t)}] = \mbox{constant}.
\end{equation}
Note that $\mean{q(t)}$ is a positive martingale bounded  by 1.
From the continuous-time martingale convergence theorem \cite{karatzas}
it follows that $\mean{q(t)}$ converges to a stable distribution as $t\to \infty$,
not necessarily a constant.

\subsubsection{The large $n$ limit}
When $n$ is large $n_k^{-1/2}$ is small, so that we can neglect
the fluctuation term in Eq.~({\ref{dynamics}) and write
\begin{equation}
\label{mean_dynamics} \frac{dq_k}{dt} = \mean q-q_k.
\end{equation}
This amounts to a mean-field approximation. We divide the nodes
into different groups according to their degrees, so that all
nodes in the same group have the same degree. If when $n$ is large
the size $n_k$ of each group is also large, then we can
approximately neglect the fluctuations within each group and
replace the group-wise random variables $m_k$, $q_k$ by their mean
values. In this sense Eq.~(\ref{mean_dynamics}) can be regarded as
a set of normal differential equations which contain deterministic
variables only.

Since $\mean q$ is now deterministic, Eq.~(\ref{eq_constant}) becomes
\begin{equation}
\label{q_constant} \mean q = \mbox{constant}.
\end{equation}
Thus Eq.~(\ref{mean_dynamics}) can be easily solved. The solution is
\begin{equation}
\label{solution_qk} q_k(t) = q_k(0) e^{-t} + \mean{q(0)} (1-e^{-t}).
\end{equation}
We see that for each $k$,
\begin{equation}
\label{limit_qk} \lim_{t\to\infty} q_k(t) = \mean q.
\end{equation}

Because $q = \sum n_k q_k / \sum n_k$ is a simple average over $q_k$, we have from Eq.~(\ref{solution_qk})
\begin{equation}
\label{solution_q} q(t) = q(0) e^{-t} + q(0) (1-e^{-t})
\end{equation}
and
\begin{equation}
\label{limit_q} \lim_{t\to\infty} q(t) = \mean q.
\end{equation}

\subsection{Interpretation of the solution}
A direct corollary of Eq.~(\ref{eq_constant}) is that if one
starts with a nontrivial initial distribution of opinions (i.e.,
the nodes are not all black or all white), then on average no
opinion will dominate in the end. This rather surprising result
was tested in a computer experiment described in Section
\ref{sec_simulation}.

In general the overall fraction of black nodes $q$ is not equal to
$\mean q$, so it can change with time. Eq.~(\ref{limit_q}) shows
that $q$ approaches $\mean q$ as time goes on. To put it more
clearly, suppose at $t=0$ the network is colored in some way such
that $q \ne \mean q$, then averagely speaking, as time passes $\mean q$ stays at its
initial value, while $q$ keeps moving towards $\mean q$. This is
also confirmed by simulation.

To better compare $q$ and $\mean q$ we rewrite their definitions as
\begin{equation}
q = \frac mn = \frac {\sum m_k}{\sum n_k};
\end{equation}
\begin{equation}
\label{weighted_average2}\mean q = \frac {\sum k p_k q_k} {\sum k
p_k} = \frac {\sum k n_k q_k} {\sum k n_k} = \frac {\sum k m_k}
{\sum k n_k}.
\end{equation}
It becomes clear that in the weighted average $\mean q$, each node
is given a weight $k$ equal to its degree. Thus,
Eq.~(\ref{limit_q}) and (\ref{weighted_average2}) says that a
high-degree node contributes more to the final fraction of colors
(decisions) than a low-degree node. Quantitatively, \emph{the
contribution of every node is proportional to its degree}. In
other words, \emph{high-degree nodes are more influential}. This
explains why a relatively small number of people with high social
ranks can affect a significant proportion of the whole society in
their decision making.

We emphasize that our theory \emph{explains, rather than assumes}
why high-rank nodes are more influential in affecting opinion
formation than low rank nodes. In fact, in our model when a node
updates its color, it puts equal weight on all its neighbors. The
chance that it will get the color from a high-degree neighbor and
the chance that it will get from a low-degree neighbor are the
same. However, statistically speaking there are more nodes in the
network that are affected by any high-degree node. In other
words, people with higher social rank are more influential because
\emph{more people pay attention to them}. Notice that this not the
same as ascribing a higher weight to the single opinion of a
high-rank member of the group.

Furthermore the fragility of opinion formation that our theory
exhibits stems from the possibility that a relatively small number
of nodes contribute a significant proportion to the weighted
$\mean q$, thus changing the whole network dramatically. This
effect was also tested by computer simulations which we will show in the next section.

\subsection{\label{sec_simulation}Computer simulations}

\subsubsection{Network creation}

The results derived in the previous sections apply to arbitrary
degree distributions. However, in order to stress the degree
effect, we performed all our simulations on a connected power-law
network of size $n=10^4$ and $\alpha=2.7$, whose (continuous)
degree distribution is given by $p_k = (\alpha-1) k^{-\alpha}$,
$k\ge 1$. A sample degree distribution for such a network is shown
in Fig.~\ref{fig_deg_dist}.

\subsubsection{\label{sec_free_rand_net}Random colored network}

We first created a random network as described and randomly
assigned 70\% of the nodes to be black and 30\% to be white. We
then randomly picked one node in the network and randomly updated
its color to be the color of one of its neighbors. This
``pick-and-update'' step was repeated $10^6$ times so that on
average each node got updated 100 times, which is a rather large
number for a network of this size. These $10^6$ steps constitute a
``sample path'' of the stochastic process, along which both $q$
and $\mean q$ were calculated as functions of $t$.

\begin{figure}
  \centering\epsfig{file=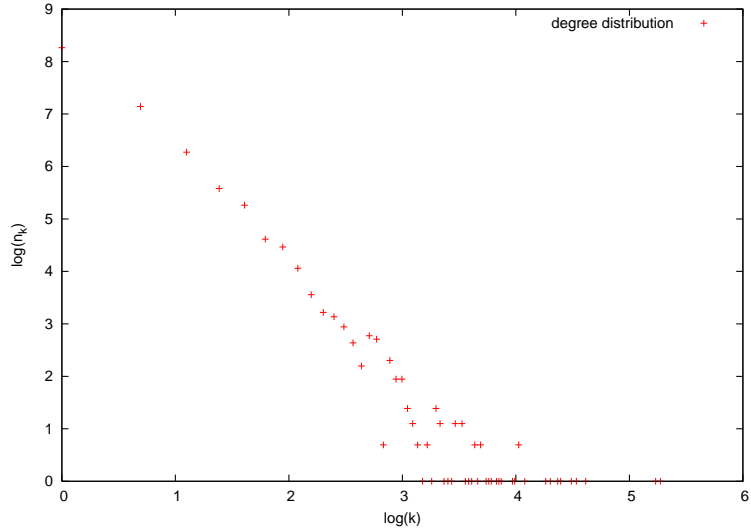, scale=.8}
  \caption{\label{fig_deg_dist}Degree distribution of a network with size $10^4$ and $\alpha=2.7$.}
\end{figure}

We repeated this experiment 100 times, each time on regenerated
networks, so that 100 sample paths were collected. Three of those
sample paths are shown in Fig.~\ref{fig_q_random} and
\ref{fig_wq_random}. As can be seen from the figures, $\mean q$
has a larger variance than $q$.

\begin{figure}
  \centering\epsfig{file=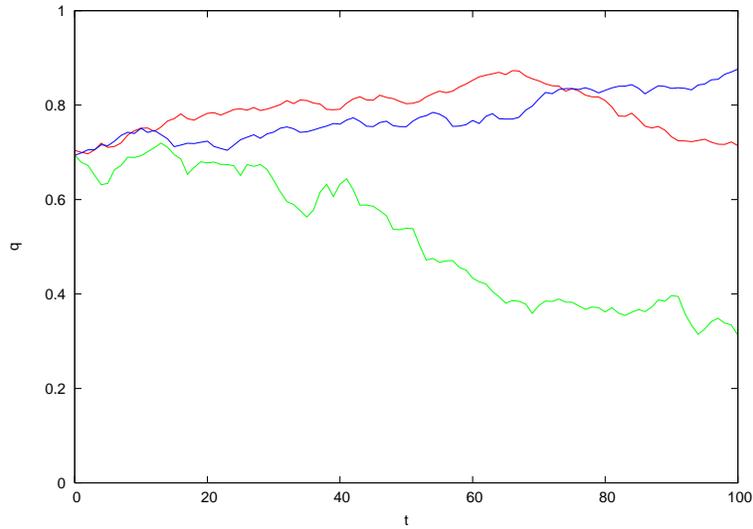, scale=.8}
  \caption{\label{fig_q_random}Evolution of the fraction of black nodes, $q$, on a free random network. The unit of time, $t$, is $10^4$ rounds. The three fraction curves are calculated along three different sample paths, each path sampled on a distinct network. As can be seen none of the three curves reaches 0 or 1 after $100 \times 10^4$ rounds, suggesting a nontrivial limit distribution.}
\end{figure}
\begin{figure}
  \centering\epsfig{file=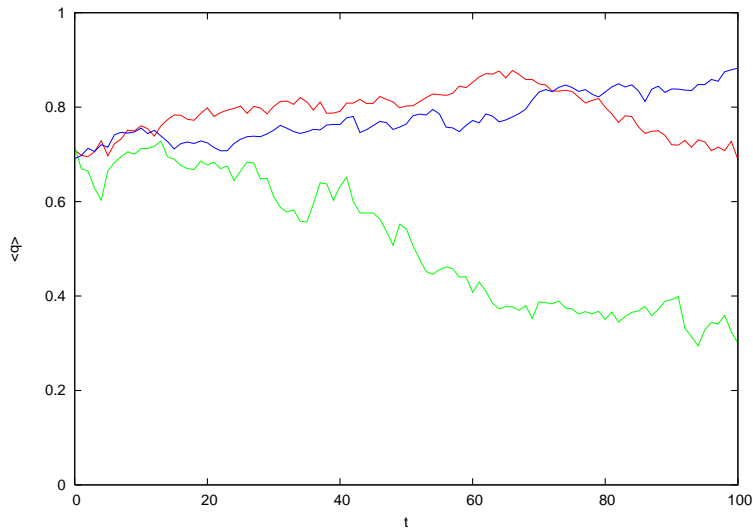, scale=.8}
  \caption{\label{fig_wq_random}Evolution of the weighted fraction of black nodes, $\mean q$, on a free random network. The unit of time, $t$, is $10^4$ rounds. The three weighted fraction curves are calculated along the same three sample paths as in Fig.~\ref{fig_q_random}.}
\end{figure}

If we take the average of $q(t)$ and the $\mean{q(t)}$ over all
100 sample paths we get estimates for $Eq(t)$ and $E\mean{q(t)}$.
These are shown in Fig.~\ref{fig_mean_random}. It is clear that
both $Eq$ and $E\mean q$ do not change with time, which confirms
the prediction of a martingale in Eq.~(\ref{eq_constant}).

\begin{figure}
  \centering\epsfig{file=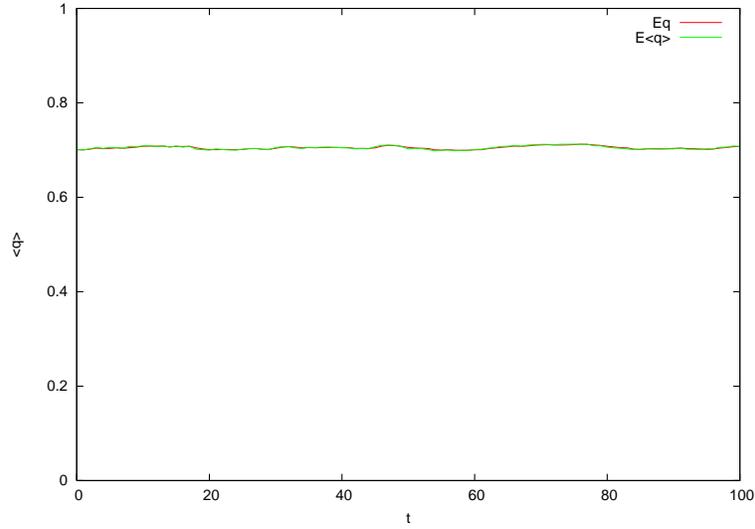, scale=.8}
  \caption{\label{fig_mean_random}The expected fraction of black nodes (red line) and the expected weighted fraction of black nodes (green line) do not change with time. The expectations are estimated by averaging over 100 sample paths.}
\end{figure}

\subsubsection{\label{sec_nonrandom_color}Nonrandom color modification}

To show that a significant proportion of nodes can be affected by
a small number of high-degree nodes, we performed the following
experiment. As in Section \ref{sec_free_rand_net}, we first
created a random network, and then randomly assigned 70\% of its
nodes to be black and 30\% to be white. \emph{We then manually
assigned the 100 highest-degree nodes to the color white.} Because
these 100 nodes constitute only 1\% of the whole network and some
of them were originally white before the manual assignment, only
less than 1\% proportion of the network is affected. In other
words, the change of $q$ due to the manual step was less than 1\%,
which can be neglected. On the other hand, because the 100
high-degree nodes contribute a significant weight to the weighted
average, the change in the value of $\mean q$ is significant and
cannot be neglected. In fact, $\mean q$ was lowered from 0.7 to
about 0.55 by the color modification.

The rest steps remain the same as in Section
\ref{sec_free_rand_net}. We again collected 100 sample paths,
three of which are shown in Fig.~\ref{fig_q_setwhite} and
\ref{fig_wq_setwhite}.

\begin{figure}
  \centering\epsfig{file=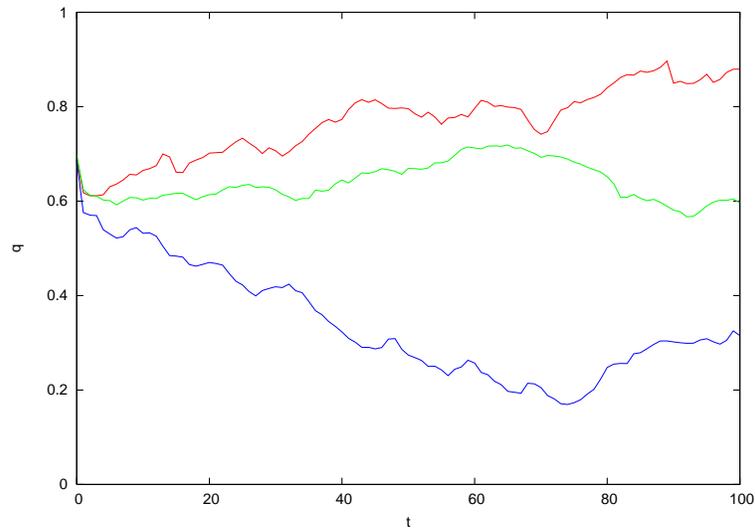, scale=.8}
  \caption{\label{fig_q_setwhite}Evolution of the fraction of black nodes, $q$, on a free network with the 100 highest-degree nodes set to white. The unit of time is $10^4$ rounds. The three fraction curves are again calculated along three sample paths on three distinct networks.}
\end{figure}
\begin{figure}
  \centering\epsfig{file=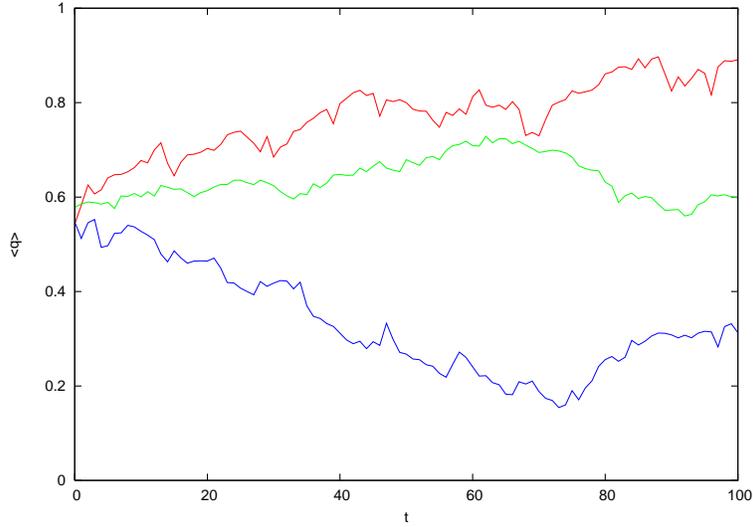, scale=.8}
  \caption{\label{fig_wq_setwhite}Evolution of the weighted fraction of black nodes, $\mean q$, on a free network with the 100 highest-degree nodes set to white. The unit of time is $10^4$ rounds. The three fraction curves are calculated along the same three sample paths as in Fig.~\ref{fig_q_setwhite}.}
\end{figure}
\begin{figure}
  \centering\epsfig{file=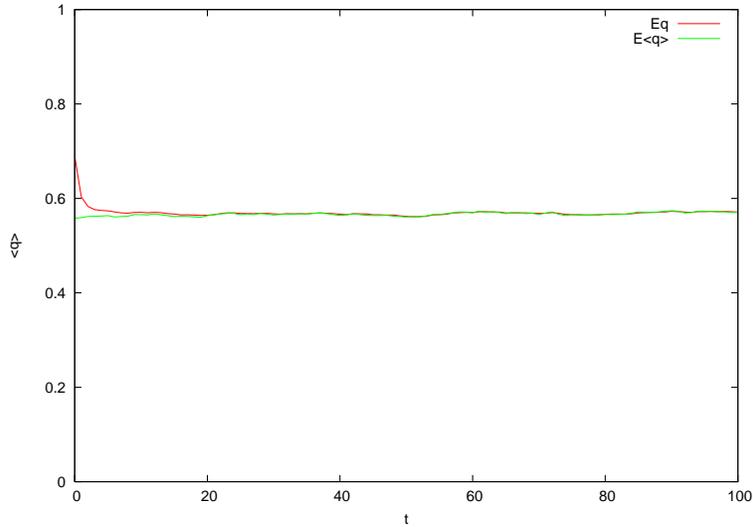, scale=.8}
  \caption{\label{fig_mean_setwhite}Evolution of the expected value of the fraction of black nodes, $Eq$, towards the expected weighted fraction $E\mean q$ as a function of time. At the beginning $Eq=0.7$ and $E\mean q=0.55$. The equilibrium $Eq = E\mean q=0.55$ is reached after about $10 \times 10^4$ rounds, i.e., after each node updates its color 10 times on average.}
\end{figure}

We also take the sample averages of $q$ and $\mean q$ and plot
them as functions of time (Fig.~\ref{fig_mean_setwhite}). It can
be seen  that $E\mean q$ again does not change with time, which
further confirms Eq.~(\ref{eq_constant}). It is also seen that
$Eq$ approaches $E\mean q$ as time goes on, as predicted by
Eq.~(\ref{limit_q}).

\section{Social networks with a sprinkle of fixed opinions}

\subsection{Dynamical equations}

So far we have assumed that the network is free in the sense that
every person-node can change its color at will any number of
times. We now extend our model to allow a fraction of the people
to have fixed opinions, which translates into nodes with fixed
colors. These recalcitrant people or nodes can be regarded as
``sources'' of the network, in the sense that they can affect
others but they themselves cannot be affected by the opinion of
others. In a social context, these nodes correspond to ``decided''
people while the other nodes correspond to ``undecided'' people.

Let $b_k$ be the proportion of degree-$k$ nodes that stay black
forever, and let $w_k$ be the proportion of degree-$k$ nodes that
stay white forever. The remaining $1-b_k-w_k$ proportion of
degree-$k$ nodes are free to change their colors as before. We now
study what the final outcome is going to be for this more
realistic case.

The difference between a free network and a network with sources
is that in the latter case when we randomly choose a node to
update, we have to make sure it is free and thus can be updated.
Suppose a degree-$k$ node is chosen. At the moment it is chosen,
there are $n_k (1-q_k)$ white nodes with degree $k$, among which
$n_k w_k$ are not free. Therefore the probability that a free
white node is chosen is
\begin{equation}
\frac{n_k(1-q_k) - n_k w_k}{n_k} = 1-q_k-w_k.
\end{equation}
Hence we need to replace $1-q_k$ by $1-q_k-w_k$ in Eq.~(\ref{prob_w_b}) to obtain
\begin{equation}
P_{w\to b}(k) = p_k (1-q_k-w_k) \mean q,
\end{equation}
Similarly, Eq.~(\ref{prob_b_w}) is modified to
\begin{equation}
P_{b\to w}(k) = p_k (q_k-b_k) (1-\mean q).
\end{equation}

Repeating the steps in the previous section, we can  reach a set
of dynamical equations similar to Eq.~(\ref{dynamics}):
\begin{equation}
\label{dynamics_source} dq_k = [\mean q-q_k+b_k(1-\mean q)-w_k\mean q] dt + \frac 1{\sqrt{n_k}} \sigma_k dB_t^{(k)},
\end{equation}
where $\sigma_k$ is a complicated function of $q_k$ which we do
not write out. When $b_k=w_k=0$ Eq.~(\ref{dynamics_source})
becomes Eq.~(\ref{dynamics}).

\subsection{The solution}

Taking the weighted average on both sides of
Eq.~(\ref{dynamics_source}), we have
\begin{equation}
d\mean q = [b_k(1-\mean q) - w_k \mean q] dt + \mean{\frac 1{\sqrt{n_k}} \sigma_k dB_t^{(k)}}.
\end{equation}
Hence $\mean q$ is no longer a martingale. If we again apply the
mean-field approximation to neglect the fluctuation terms, we get
\begin{equation}
\frac{d\mean q}{dt} = \mean b(1-\mean q) - \mean w\mean q.
\end{equation}
The equilibrium condition is obtained by setting the right hand side equal to zero ($q_\infty=q(t=\infty)$):
\begin{equation}
\mean b(1-\mean{q_\infty}) - \mean w\mean{q_\infty} = 0,
\end{equation}
which gives
\begin{equation}
\label{equilibrium_q}\mean{q_\infty} = \frac{\mean b}{\mean b+\mean w}.
\end{equation}
Therefore as $t\to\infty$, $\mean{q(t)}$ converges to a fixed
fraction equal to the weighted proportion of non-free black nodes
among all non-free nodes. We see that the final proportion does
not depend on the random initial assignment of the colors of the
free nodes, although it is possible that the convergence needs
such a long time that it can never be reached in reality. Anyway,
Eq.~(\ref{equilibrium_q}) shows that the weighted average again
plays an important role, indicating that high-degree nodes are
more influential to the final outcome.

\section{The effect of undecided individuals}

\subsection{Model}
In the first two models  we assumed that each person or node can
make decisions repeatedly for any number of times. However, in
some circumstances, once a node makes a decision it remains
unchanged during the whole process of opinion formation.
Accordingly, we will now assume that there are two kinds of people
or nodes, decided and undecided. A decided node has opinion either
black or white, which does not change with time, while an
undecided node has no color at the beginning but can obtain one
from one of his neighbors after an update of its state. Once it
gets a color, it becomes decided and its color stays fixed
forever. To conclude, each node has three possible states: black,
white and undecided.

As before, at each step we randomly pick a node from the network
and check its state. If it already has a color (decided), we do
nothing. If it is undecided, we randomly pick one of its neighbor.
If that neighbor is also undecided, we again do nothing, otherwise
we update the first node's color to be the same as its neighbor's.

\subsection{Solution}

Let $b_k$ and $w_k$ be the proportion of black and white nodes in
the network, respectively.  e assume that $b_k+w_k<1$ at $t=0$ so
that there are a finite number of undecided nodes at the
beginning.

We calculate the probability that the number of $k$-degree black
nodes will be increased by one during an update. For this to
happen, first we have to choose an undecided node in step 1, which
happens with probability $1-b_k-w_k$, and then its neighbor we
choose in step 2 has to be black, which happens with probability
$\sum kp_k b_k/\sum kp_k$. Thus we have (again neglecting the
fluctuation term by mean-field approximation)
\begin{equation}
\label{three_state_dynamics1}\frac{db_k}{dt} = (1-b_k-w_k) \mean b,
\end{equation}
and similarly
\begin{equation}
\label{three_state_dynamics2}\frac{dw_k}{dt} = (1-b_k-w_k) \mean w.
\end{equation}
Eq.~(\ref{three_state_dynamics1}) and
(\ref{three_state_dynamics2}) govern the dynamics of the system.

Taking the weighted average of Eq.~(\ref{three_state_dynamics1}) and (\ref{three_state_dynamics2}), we obtain
\begin{equation}
\label{three_state_weighted1}\frac{d\mean b}{dt} = (1-\mean b-\mean w) \mean b,
\end{equation}
and
\begin{equation}
\label{three_state_weighted2}\frac{d\mean w}{dt} = (1-\mean b-\mean w) \mean w.
\end{equation}
To solve Eq.~(\ref{three_state_weighted1}) and
(\ref{three_state_weighted1}), we take their sum and define
$f=1-\mean b-\mean w$ to get
\begin{equation}
\frac{df}{dt} = f(1-f).
\end{equation}
Now $f$ can be solve as
\begin{equation}
f(t) = \frac{1-f_0}{f_0 e^t + 1-f_0},
\end{equation}
where $f_0=f(0)=1-\mean{b(0)}-\mean{w(0)}$. Putting this back into
Eq.~(\ref{three_state_weighted1}) and
(\ref{three_state_weighted2}), we can solve out $\mean b$ and
$\mean w$, which we write down here:
\begin{equation}
\mean b = \frac{\mean{b(0)} e^t}{f_0 e^t + 1-f_0}, \quad \mean w = \frac{\mean{w(0)} e^t}{f_0 e^t + 1-f_0}.
\end{equation}
Hence
\begin{equation}
\frac{\mean{b(t)}}{\mean{w(t)}} = \frac{\mean{b(0)}}{\mean{w(0)}} = \mbox{const}.
\end{equation}
We see that the weighted black-to-white ratio does not change with
time. In fact, this can be seen from
Eq.~(\ref{three_state_weighted1}) and
(\ref{three_state_weighted2}) directly, where the increments of
$\mean b$ and $\mean w$ is proportional to $\mean b$ and $\mean
w$, respectively.

\section{Remarks}

\subsection{Information asymmetries}
Since our model makes no assumption about the degree distribution,
it applies to all kinds of networks including power-law networks
and exponential networks. Furthermore, we can further extend our
model to describe informational asymmetries in such a way  that it
is possible for $A$ to get information from $B$ but $B$ cannot get
information from $A$. This corresponds to the study of our model
on a directed graph and is illustrated in
Fig.~\ref{directed_graph}. In this example  $B,C,D,E$ can get
information from $A$ but $A$ can only get information from $D$. A
directed graph resembles more closely a real life social network,
in which low-rank people pay more attention to high-rank people
than the other way around.

To generalize our model for undirected graphs, from the point of
view of our notation we need to do is to replace the numerous
appearances of ``degree'' by ``outgoing degree'' in Table
\ref{tab_symbols}. As an  example, $p_k$ now stands for ``outgoing
degree distribution''. We point out that the outgoing degree
distribution of a directed graph can be very different from the
degree distribution of the same graph viewed as an undirected
graph. For example, node $D$ in Fig.~\ref{tab_symbols} has
outgoing degree 1 as a directed graph but degree 3 as an
undirected graph.

Under the new definition, all our previous results still hold.

\begin{figure}
  \centering\epsfig{file=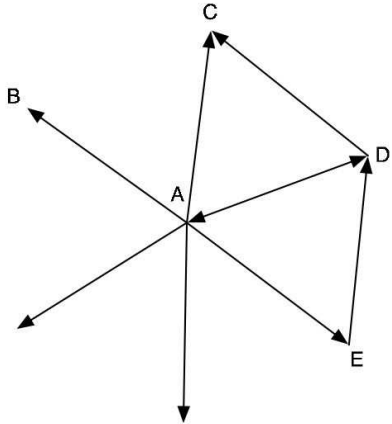, scale=.65}
  \caption{\label{directed_graph}A directed graph. The arrow marks the direction of information flow. For example,
  $A$ points to $C$ means $C$ can get information from $A$ but not the other way.}
\end{figure}

\section{Discussion}

In this paper we presented a theory of opinion formation that
explicitly takes into account the structure of the social network
in which individuals are embedded. The theory assumes asynchronous
choices by individuals among two or three opinions and it predicts
the time evolution of the set of opinions from any arbitrary
initial condition. We showed that under very general conditions a
martingale property ensues, i.e. the expected weighted fraction of
the population that holds a given opinion is constant in time. By
weighted fraction we mean the fraction of individuals holding a
given opinion, averaged over their social connectivity (degree).\footnote{Note that in the context of epidemic control Dezso and Barabasi
established a similar result that it is more efficient to cure the
high degree nodes first \cite{dezso, dodds}. However, they did not
give a quantitative definition of importance like our proportional
relation, nor did they propose any convergence law.}
Most importantly, this weighted fraction is not either zero or
one, but corresponds to a non-trivial distribution in the long
time limit. This coexistence of opinions within a social network
is in agreement with the often observed locality effect, in which
an opinion or a fad is localized to given groups without infecting
the \emph{whole} society.

Our theory further predicts that a relatively small number of
individuals with high social ranks can have a larger effect on
opinion formation than individuals with low rank. By high rank we
mean people with a large number of social connections. This
explains naturally a fragility phenomenon frequently noted within
societies, whereby an opinion that seems to be held by a rather
large group of people can become nearly extinct in a very short
time, a mechanism that is at the heart of fads.

These predictions were verified by computer experiments and
extended to the case when some individuals hold fixed opinions
throughout the dynamical process. Furthermore, we dealt with the
case of information asymmetries, which are characterized by the
fact that some individuals are often influenced by other people's
opinions while being unable to reciprocate and change their
counterpart's views.

While the assumption of only two or three opinions within a social
network may seem restrictive, there are many real world instances
where people basically choose among points of view. Examples are
elections in two party systems, management fads which consultants
and executives need to decide whether to implement or not, and
highly polarized attitudes towards government actions in many
social settings. Our finding that social structure and ranking do
affect the formation of these opinions and that they can coexist
with each other are in agreement with many empirical observations.

Our findings also cast doubt on the applicability of tipping
models to a number of consumer behaviors \cite{gladwell}. While
there are clear thresholds in the spread of innovations when
network externalities are at play \cite{rogers,loch} it is not
clear that the same phenomenon is observed in situations where
externalities are not at play. In most of the consumer behaviors
that have been ``explained'' by tipping point ideas one still
observes the coexistence of the old and the new preference or
opinions over long times, in contrast with the sudden onset seen
in the case of positive externalities.

\bigskip We thank Lada Adamic and Chenyang Wang for some useful suggestions.

\end{document}